# Stairs Detection for Enhancing Wheelchair Capabilities Based on Radar Sensors

Sherif Abdulatif*, Bernhard Kleiner*, Fady Aziz*, Christopher Riehs*, Rory Cooper†, Urs Schneider*

*Department of Biomechatronic System, Fraunhofer Institute for Manufacturing Engineering and Automation IPA

†Human Engineering Research Laboratories, University of Pittsburg

Email: {sherif.abdulatif, bernhard.kleiner, fady.aziz, christopher.riehs, urs.schneider}@ipa.fraunhofer.de \rcooper@pitt.edu

*Abstract*—Powered wheelchair users encounter barriers to their mobility everyday. Entering a building with non barrier-free areas can massively impact a persons mobility related activities. There are a few commercial devices and some experimental that can climb stairs using for instance adaptive wheels with joints or caterpillar drive. These systems rely on the use for sensing and control. For safe automated obstacle crossing, a robust and environment invariant detection of the surrounding is necessary. Radar may prove to be a suitable sensor for its capability to handle harsh outdoor environmental conditions. In this paper, we introduce a mirror based two dimensional Frequency-Modulated Continuous-Wave (FMCW) radar scanner for stair detection. A radar image based stair dimensioning approach is presented and tested under laboratory and realistic conditions.

## I. INTRODUCTION

Mobility is a key factor in determining autonomy and independence for people with severe disabilities (PsD). Electric powered wheelchairs (EPW) are an important means of providing mobility to PsD across the age span. Tips and falls are the most frequent reason that PsD who use EPW report to emergency rooms [1]. There is a need for more capable EPW that can reduce the risk of tips/falls or loss of control, which occur in 30%-65% of users each year [2]. Tips and falls related to EPW crashes have a significant effect on PsD. Improvements have been shown in EPWs in the past 20 years [3], [4], including reliability, better suspension to minimize vibration exposure and expanded user interfaces. Despite some improvements, current EPW design limits most users to drive in indoor environments and outdoors with mostly flat barrier-free environments. Furthermore, PsD using EPW have difficulties and thus often avoid, driving over uneven terrain or overcoming architectural barriers such as curbs and terrains non-compliant to accessibility standards [5].

Many studies were done to allow either manual or powered wheelchairs to climb the stairs. One solution introduced for stair climbing problem is installing tracks to wheelchairs [6]. Such approach needs no prior knowledge about the stairs, but irregular stair edges can easily cause slipping since all weight of the wheelchair rests on the stair edges. Another safer edge independent solution is leg-based wheelchairs to climb stairs as step by step or even in a high single step according to the legs elevation capability [7]. For this approach a prior knowledge about stairs dimensions such as depth and height is crucial for safety. In [8], [9], a LIDAR and camera based more general surrounding detection techniques were used to identify objects for autonomous driving wheelchairs. However, both sensors suffer many limitations due to different lighting, surface material and harsh outdoor environmental conditions.

## II. STAIR DETECTION SETUP

### A. FMCW-Radar

Radar is introduced in this paper as an efficient tool for stair detection for its capability to handle outdoor measurement conditions like sun light, dust and unstructured terrain. The radar used in this paper is a compact 94 GHz FMCW radar with adjustable parameters and an aperture of about 11° [10]. The used signal modulation on the FMCW radar is a continuous sequence of chirps and each chirp is defined as a linear increasing frequency signal over time. As shown in Fig. 1, the received echo signal will have a frequency shift ($f_D$) corresponding to the velocity information and a time shift ($\Delta t$) corresponding to the range information.

The main adjustable parameters on the modulation related to our application are the range resolution and the maximum detectable range. The range resolution ($r_{res}$) is defined as the minimum distance between targets where the radar is still able to distinguish them as multiple targets. Based on Eq. 1, the range resolution is only a function of the radar bandwidth ($B$) and is set to 1.5 cm (lower than the minimum possible distance between two consecutive steps) based on a bandwidth of 10 GHz. The maximum range ($r_{max}$) is defined as the maximum distance from the radar where a target is still detected. From Eq. 2, the maximum range is just the range resolution relation scaled with the number of samples per chirp ($N_s$). Given the bandwidth is set to 10 GHz, then $N_s$ is set to 200 samples per chirp to get a maximum range of 3 m which is suitable for our application.

$$r_{res} = \frac{c}{2B} \quad (1)$$

$$r_{max} = \frac{cN_s}{2B} \quad (2)$$

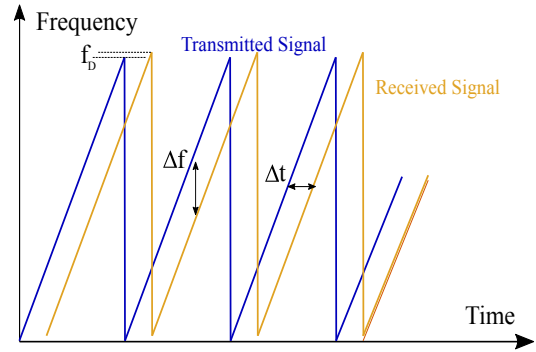

Fig. 1: FMCW Radar transmitted and received chirps.

To extract the range information of the targets in the radar scene, a Fast Fourier Transform (FFT) is applied to each ramp in the down converted received signal. Accordingly, the frequency spectrum will have a strong DC component that can be eliminated by a high pass filter. In addition to peaks at frequency shifts corresponding to the time delays ($\Delta t$). As shown in Fig. 2, the frequency of each peak represents a range of a certain target and the power is correlated with the Radar Cross Section (RCS) of the corresponding target. This RCS is strongly affected by the angle of radar beam incidence, the size, the material and the shape of the target [11]. Targets with sharp edges are known to show significant high power reflections, thus radar is a convenient sensor for stairs detection and dimensioning.

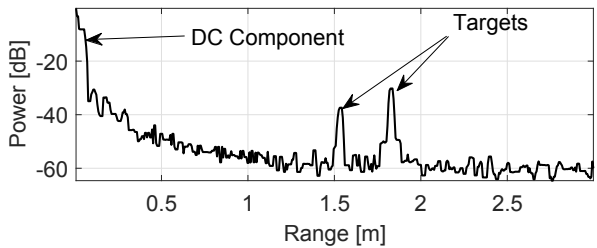

Fig. 2: Range profiles with peaks at the corresponding targets.

### B. Rotating Mirror Based Scanner

A Single Input Single Output (SISO) radar with one transmitting and receiving antenna can only detect range of objects within its beam aperture without angular information. Thus, to get a precise scanning of the scene, a right aperture radar is needed with an angular resolution capability. Scanning is done based on a mechanical motion of the radar and acquiring range profiles at each point, then mapping range data to correct radar positions. Different scanning techniques are often used as applying a translational motion to the radar over the height-dimension or rotating the radar over the depth plane [12]. These scanning technique are rather simple solutions, but not compact enough for wheelchair application.

In this paper, we introduce a compact solution for scanning by keeping both the radar height and orientation constant and rotating the radar beam only based on a rotating mirror. The designed mirror is concave and the surface is made from aluminum such that at any instant the radar beam is reflected by 90° to allow scanning in the sagittal plane (height and depth) as shown in Fig. 3. The center of the mirror is horizontally aligned with the radar lens and placed at a distance of 22 cm. The rotating scanner structure is placed horizontally at a suitable wheelchair height of 40 cm. The mirror is designed with a 5° aperture and this reduction from the original radar beam aperture (11°) will increase the vertical resolution capability. The mirror is mounted to a small shaft and can be freely moved over 340° angular range. The mirror rotation is controlled by a step motor such that at each measurement instant the mirror is static.

In our setup shown in Fig. 3, the mirror is rotated by an angular resolution of $\theta_{res}$ for each range spectrum measurement. This angular resolution strongly affects the height resolution (the translational distance moved in the height plane based on this angular rotation). Based on Eq. 3, the height resolution ($h_{res}$) at a distance ($d$) is directly proportional with the angular resolution ($\theta_{res}$). Considering a stair detected at the chosen radar maximum range of 3 m, an angular resolution ($\theta_{res}$) of 0.25° is chosen to satisfy a height resolution ($h_{res}$) of 1 cm which is enough for our application.

$$h_{res} = d \tan \theta_{res} \quad (3)$$

Based on experiments on different staircases, the angular dynamic range is chosen from -20° above the horizontal depth plane to 50° below the horizontal depth plane. Within the specified angular range, each range spectrum is mapped to a world coordinate system. Finally, a 2D intensity map in the sagittal plane is generated representing reflected signals from different objects as shown in Fig. 4. At a distance of 1 m three steps are detected on a wooden stair as three vertical high intensity planes at different heights and depths.

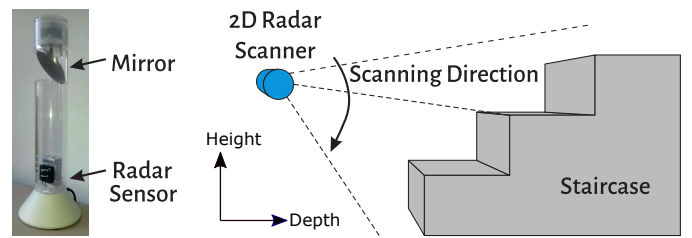

Fig. 3: 2D scanner (left) and the experimental setup (right).

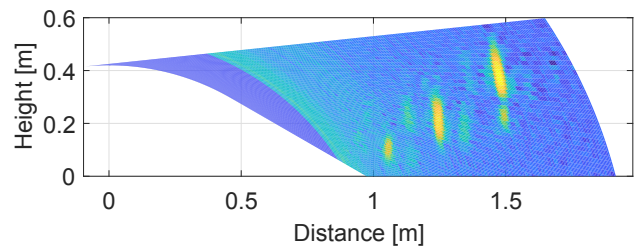

Fig. 4: 2D Radar scanned image of a 3 steps staircase.

### III. STAIR DETECTION ALGORITHM

In this part, we present a particle filter based plain detection algorithm to identify high intensity parts in the staircase scanned image shown in Fig. 4. The idea behind the particle filter is to represent a probability distribution function (pdf) based on random samples with corresponding weights. Thus multiple objects will be represented as a pdf. The algorithm will work by initializing particles all over the image and applying iterative resampling to initialized particles till the particles converge to high power areas in the intensity map. Based on the resampled particles distribution, clustering is applied to separate detected stairs. Furthermore, particles distribution in each cluster will be used to estimate height and depth of each detected stair.

### A. Particles Initialization and Resampling

First part of the plain detection algorithm will be the particle filter initialization phase. In this phase $N$ particles are initialized with a pre-defined distribution. As shown in Fig. 5, each particle $s_i$ (blue point) is represented as:

$$s_i = (x_i, y_i, p_i) \quad (4)$$

where $x_i, y_i$ represents the position of the $i^{th}$ particle (distance, height) in the sagittal plane, $p_i$ represents the power at this particular position and $i$ represents the particle index from 1 to $N$. After initialization, all the powers are normalized to weights and each particle will get a weight $w_i$ such that the sum of all weights ($\sum_{i=1}^{N} w_i$) is equal to 1.

In the second phase, resampling by replacement based on the weights $w_i$ is applied to the initial state particles distribution. During resampling, particles with high weights are more likely to be selected multiple times and replace particles with low weights [13]. Resampling is repeated recursively for multiple amount of times to insure convergence to high intensity plains (red points) as shown in Fig. 5.

Particles distribution during the initialization phase can strongly affect how fast particles can converge to the high intensity areas during resampling phase. One common approach for particles initialization is to uniformly distribute all particles over areas with defined power values in the image as shown in Fig. 5. This approach is rather simple, but it can take about 15 iterations to converge to the correct high intensities.

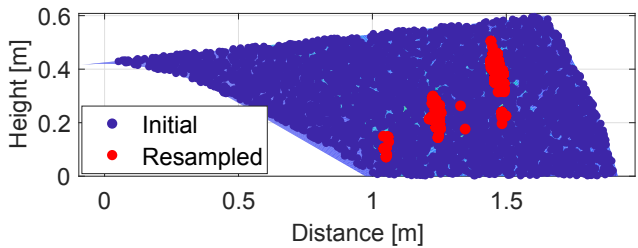

Fig. 5: Uniform particles distribution over scanned image.

In this paper, we use a Gaussian Multi-Modal (GMM) distribution initialization technique where $M$ high intensity locations are randomly chosen within the image. A subset $N/M$ (mode) of the particles is then initialized with a suitable variance and a mean equal to the corresponding high intensity position. The same approach is applied till we initialize all required $N$ particles as shown in Fig. 6. After initialization, weight normalization and resampling is applied to each subset alone which will allow faster convergence to high intensity planes [14]. Furthermore, this distributed weight normalization and resampling approach will insure convergence to all high intensities in the scan even if some areas are lower in power than others. The used multi-modal Gaussian initialization was tested on different staircase scans and it can converge to high intensity areas within 5 resampling iterations. For the sake of lowering the complexity, number of particles $N$ is chosen to be 1000 particles and number of modes $M$ as 10 modes. Finally, to reduce the overall complexity before applying clustering to the resampled particles, redundant particles with the same exact positions and powers due to resampling with replacement are removed.

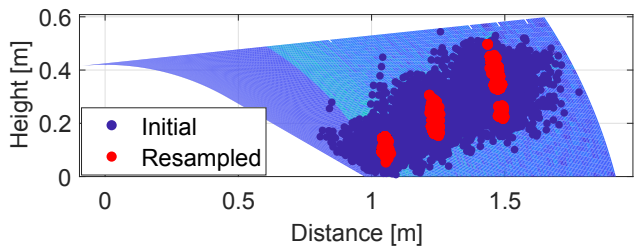

Fig. 6: GMM particles distributions over scanned image.

## B. Clustering and Rejection

In order to separate the steps detected in an image, clustering is applied to the resampled particles based on their $(x, y)$ positions and corresponding power intensities. Accordingly, cluster bounds are generated based on an adaptable sensitivity parameter ($\rho$) to assign each particle to a cluster. The clustering sensitivity $\rho$ can have values between 0 and 1. Increasing the sensitivity will result in more bounds and separations, thus more detected clusters. In the proposed setup, the sensitivity parameter $\rho$ was tunned over different scans and is finally set as 0.5 to insure fair clustering.

As shown in Fig. 7, fair clustering can result sometimes in erroneous clusters that can be redundant (belonging to same object) or outliers clusters. Outlier clusters are mostly scattered particles which did not converge to high intensity area. Such scatters are known to have a low number of particles which can be identified and removed. A threshold on the number of particles in each cluster is introduced to identify if a cluster is an outlier. In our case, a rather simple threshold based on 10% of the mean number of particles in all clusters is used. A cluster with particles not satisfying such threshold will be removed from the detected clusters. To insure uniform distribution of particles in each cluster, an outliers detection and removal is applied to particles in each cluster.

Moreover, redundant clusters which belong to the same object must be either fused or the cluster with lower number of particles is rejected as marked in Fig. 7. Since a staircase is known to be detected, multiple clusters sharing relatively close position information in either depth or height are identified as redundant. A distance threshold of 10 cm is chosen to identify if multiple clusters are redundant, then the cluster with the lower relative position is rejected to get the final detected clusters as shown in Fig. 8.

Finally, the detected number clusters after rejection corresponds to the number of stairs the wheelchair can climb. The particles distribution in each cluster will be used for the sake of dimensioning. In the next part, implemented enhancements to our stair detection algorithm is presented in terms of faster particles convergence and adding more possible use cases.

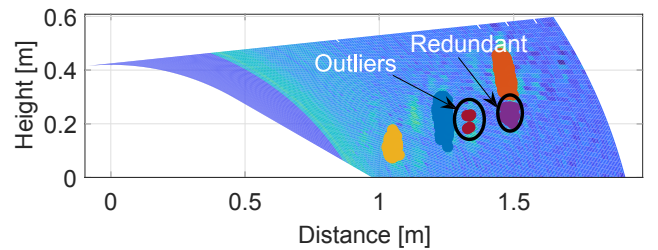

Fig. 7: Clustered particles over scanned image.

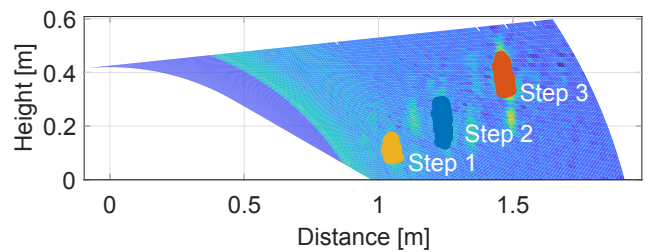

Fig. 8: Final detected clustered particles over scanned image.

## C. Clutter Filtering

During the radar scan and especially at long distances as the beam area is wider, it is likely to get reflections from unwanted objects often referred to as clutter. Due to the variability of the mixed clutter and noise, an adaptive thresholding technique for suppressing spectrum noise and keeping wanted target peaks only is needed. Constant False Alarm Rate (CFAR) is a widely used adaptive thresholding method in radar systems to separate target peaks from neighbor noise. CFAR detection is based on gain control to preserve a constant rate of false target detections in the case of varying clutter and noise levels.

There are different techniques for applying CFAR in the radar reception to prevent high false alarm rates in the presence of interference such as jamming or clutter residue. In this paper, Cell-Averaging (CA-) CFAR [15] is applied for its simplicity. In CA-CFAR, a sliding window over the range profile is used to compute the average power within the window cells. The sliding window averaging considers a potential peak to be in the middle, thus the average is computed without considering the middle cells. As shown in Fig. 9, the required target peaks exceed the CFAR threshold and otherwise is considered as noise.

To insure presence of target stairs only in radar scans, any range power below the CFAR threshold is assigned a value equal to the minimum received power and values exceeding the threshold are taken as it is. After coordinate correction at each mirror rotation, a 2D CFAR intensity map in sagittal plane can be produced. As shown in Fig. 10, the high intensity peaks representing stairs are clearly seen as high yellow values. On the other hand, noise and clutter can be identified as complete blue background. The CFAR technique was tested on different stair cases and it shows clear improvement in clutter and noise rejection. Moreover, this CFAR technique has a major influence in the particle filter complexity as particles can converge much faster to high intensity areas. As mentioned above the stair detection particle filter algorithm can take up to 5 iterations to converge to correct steps. CFAR technique was tested on several measurements and the particles can converge during resampling within 2 iterations only.

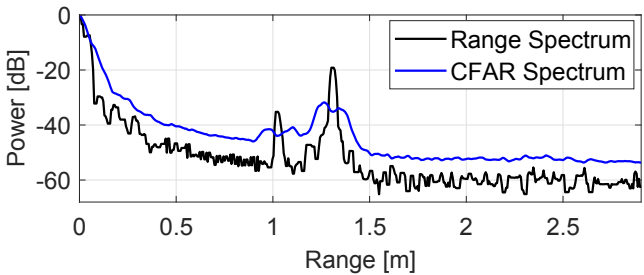

Fig. 9: CFAR threshold over detected range profiles.

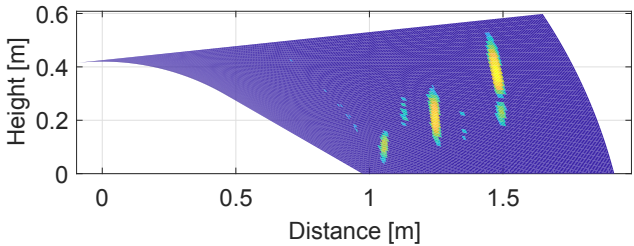

Fig. 10: 2D Radar scanned CFAR image of a 3 steps staircase.

## IV. STAIRS DIMENSIONING

After applying our proposed particle filter based stair detection algorithm and clutter filtering, now particles are correctly distributed over each step and can be used for dimensioning. For successful stair climbing, the number of stairs to climb represented as the number of detected clusters, depth and height of each step is needed.

**Depth** can be estimated by first subtracting the distance between the radar and the mirror (22 cm) mention in Section II from all the range measurements. Now the radar can be considered at 0 position and the distance between the radar and each step can be estimated based on particles distribution in the $x$-plane. This can be achieved by computing the weighted average of particles position over the $x$-plane in each cluster. Given number of particles in the $i^{th}$ cluster is $N_i$ and each particle depth position as $x_j$ and weight as $w_j$, where index $j$ vary from 1 to $N_i$. Then all the particles weights are normalized to 1 in each cluster and the depth of this cluster ($d_i$) can be estimated as:

$$d_i = \sum_1^{N_i} w_j . x_j \quad (5)$$

**Height** can be estimated based on first step appearance in the height plane. In comparison to laser, the origin of reflection with the experimental setup can only be measured directly with an uncertainty of 1.8 cm in $y$-Position at a range of 2 m. To overcome this limitation we introduce the usage of a 3 dB beam model of the radar unit. As shown in Fig. 11, the aperture is defined as the angle between upper and lower 3 dB ranges and in our method the height error is estimated as distance from beam center and the lower 3 dB beam (half aperture). Accordingly, the correct height can be estimated by getting the position of the top particle in each cluster. Then the height error subtracted from the top particle $y$-position is computed based on Eq. 3 where distance is considered as depth $d_i$ and $\theta_{res}$ as half the mirror aperture (2.5°).

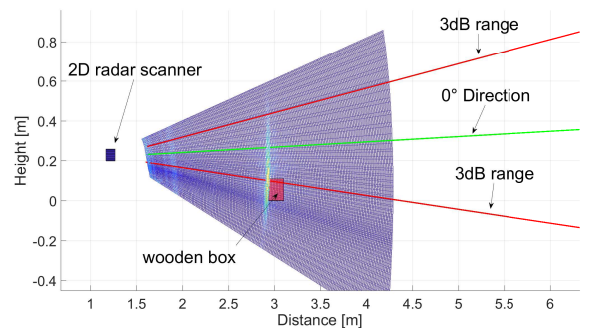

Fig. 11: Height estimation using a 3 dB beam model

## V. RESULTS

The introduced scanner was then placed over a mobile surface and was used to scan several staircases. In this part, scans of staircases with the same number of steps (3 steps) is used for fair comparison. As shown in Fig. 12, *Stair A* is a wooden constructed staircase for initial tests and *Stair B* was used to test dark ceramic material reflections. Finally, *Stair C* was used to test the algorithm detection capability of floating steps. The scanner was used to collect the scans over these staircases with the same initial position (0.5 m) and height (0.4 m). The particle filter was then used for steps detection and dimensioning, then the stairs are reconstructed and compared to real dimensions as shown in Fig. 13. The errors of both depth and height estimations of mentioned staircases are shown in Table I and detected steps are ordered in ascending depth. The depth estimation in all cases is always less than 0.5 cm and height estimation is a bit worth as it is additionally influenced by the mechanical model and aperture correction. Moreover, the height error scales with the depth as explained in Eq. 3.

Table I: Distance errors over tested staircases.

| Stairs/ Accuracy | Depth [cm] | | | Height [cm] | | |
|---|---|---|---|---|---|---|
| | Step 1 | Step 2 | Step 3 | Step 1 | Step 2 | Step 3 |
| Stair A | 0.2 | 0.4 | 0.3 | 0.5 | 0.7 | 1 |
| Stair B | 0.1 | 0.3 | 0.5 | 0.3 | 1 | 1.2 |
| Stair C | 0.2 | 0.1 | 0.4 | 0.6 | 0.6 | 1.5 |

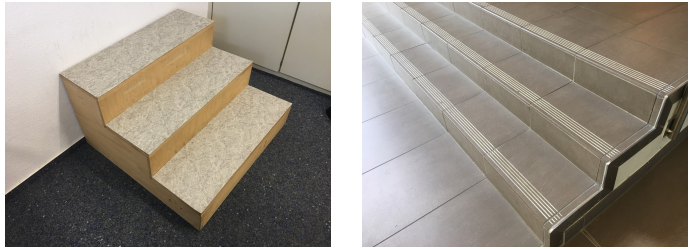

(a) Stair A  (b) Stair B

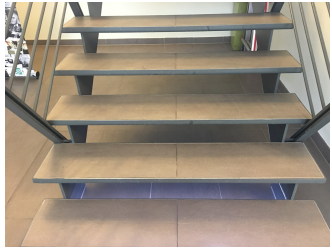

(c) Stair C

Fig. 12: Different stair types used for testing.

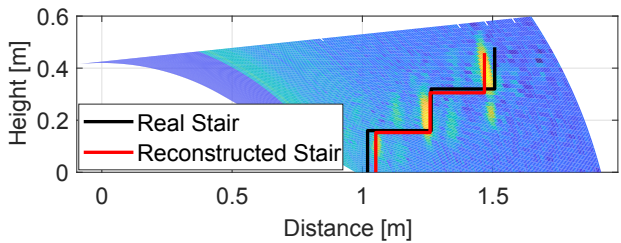

Fig. 13: Reconstructed staircase against real dimensions.

## VI. Conclusion and Further Scenarios

In this paper we introduced a mirror based stair scanner for wheelchair stair climbing applications. Moreover, we introduced a detection and dimensioning technique based on particle filter. The scanner was tested and showed good results for the desired application. We mainly addressed the scenario of a wheelchair user climbing a staircase (up stair detection), thus the radar scanner was used to scan over the depth plane to identify and dimension steps in the staircase. One other possible scenario that will yield similar results will be to apply our radar scanner in the application of downstairs detection.

In this case, the radar scanner will work by scanning over the height plane as shown in Fig. 14. This scenario is particularly achievable due to the introduced compact rotating mirror scanner. This proposed structure can provide the wheelchair user with additional capabilities of scanning in different plane directions. This can be achieved by mounting the structure to a movable pole which can move the scanner over the required staircase.

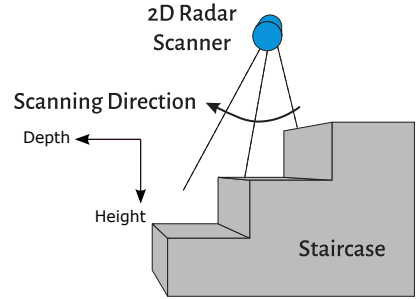

Fig. 14: Scanner illustration for downstairs detection scenario.

## References

[1] W.-Y. Chen, Y. Jang, J.-D. Wang, W.-N. Huang, C.-C. Chang, H.-F. Mao, and Y.-H. Wang, "Wheelchair-related accidents: relationship with wheelchair-using behavior in active community wheelchair users," *Archives of physical medicine and rehabilitation*, vol. 92, no. 6, pp. 892–898, 2011.

[2] H. Xiang, A. Chany, and G. A. Smith, "Wheelchair related injuries treated in us emergency departments," *Injury prevention*, vol. 12, no. 1, pp. 8–11, 2006.

[3] D. Ding and R. A. Cooper, "Electric powered wheelchairs," *IEEE Control Systems*, vol. 25, no. 2, pp. 22–34, 2005.

[4] H. Wang, H.-Y. Liu, J. Pearlman, R. Cooper, A. Jefferds, S. Connor, and R. A. Cooper, "Relationship between wheelchair durability and wheelchair type and years of test," *Disability and Rehabilitation: Assistive Technology*, vol. 5, no. 5, pp. 318–322, 2010.

[5] USA Access Board, "Americans with disabilities act and architectural barriers act accessibility guidelines," 2004.

[6] T. Chair, http://www.topchair.net, 2015.

[7] J. Candiotti, S. A. Sundaram, B. Daveler, B. Gebrosky, G. Grindle, H. Wang, and R. A. Cooper, "Kinematics and stability analysis of a novel power wheelchair when traversing architectural barriers," *Topics in Spinal Cord Injury Rehabilitation*, vol. 23, no. 2, pp. 110–119, 2017.

[8] H. Grewal, A. Matthews, R. Tea, and K. George, "Lidar-based autonomous wheelchair," in *Sensors Applications Symposium (SAS)*. IEEE, 2017, pp. 1–6.

[9] T. H. Nguyen, J. S. Nguyen, D. M. Pham, and H. T. Nguyen, "Real-time obstacle detection for an autonomous wheelchair using stereoscopic cameras," in *Engineering in Medicine and Biology Society, 29th Annual International Conference*. IEEE, 2007, pp. 4775–4778.

[10] C. Zech, A. Hülsmann, M. Schlechtweg, S. Reinold, C. Giers, B. Kleiner, L. Georgi, R. Kahle, K.-F. Becker, and O. Ambacher, "A compact w-band lfmcw radar module with high accuracy and integrated signal processing," in *European Microwave Conference (EuMC)*. IEEE, 2015, pp. 554–557.

[11] E. F. Knott, *Radar cross section measurements*. Springer Science & Business Media, 2012.

[12] J. P. Fitch, *Synthetic aperture radar*. Springer Science & Business Media, 2012.

[13] B. Efron and R. J. Tibshirani, *An introduction to the bootstrap*. CRC press, 1994.

[14] L. M. Murray, A. Lee, and P. E. Jacob, "Parallel resampling in the particle filter," *Journal of Computational and Graphical Statistics*, vol. 25, no. 3, pp. 789–805, 2016.

[15] M. A. Richards, *Fundamentals of radar signal processing*. Tata McGraw-Hill Education, 2005.